\documentclass[twocolumn]{aastex631}
\usepackage{amsmath}
\usepackage{xspace}
\usepackage{lipsum}

\newcommand{\nud}{$\Delta\nu_d$\xspace}
\newcommand{\taud}{$\tau_d$\xspace}

\newcommand{\ppcc}{$\,$pc$\,$cm$^{-3}$\xspace} 
\newcommand{\HII}{$\textrm{H}\scriptstyle\mathrm{II}$}

\begin{document}



\title{A direct measurement of the electron density turbulence parameter $C_1$ and\\implications for the emission size of the magnetar XTE J1810-197}

\author[0000-0002-4629-314X]{Visweshwar Ram Marthi}
\affiliation{National Centre for Radio Astrophysics \\
Tata Institute of Fundamental Research \\
Post Bag 3, Ganeshkhind, Pune - 411 007, INDIA}

\author[0000-0002-0862-6062]{Yogesh Maan}
\affiliation{National Centre for Radio Astrophysics \\
Tata Institute of Fundamental Research \\
Post Bag 3, Ganeshkhind, Pune - 411 007, INDIA}



\begin{abstract}
We report a direct measurement of  the electron density turbulence parameter $C_1$, enabled by 550-750~MHz baseband observations with the upgraded Giant Metrewave Radio Telescope. The parameter $C_1$ depends on the power law index of the wavenumber spectrum of electron density inhomogeneities in the ionized interstellar medium. Radio waves propagating through the inhomogeneous ionized medium suffer multipath propagation, as a result of which the pulsed emission from a neutron star undergoes scatter broadening. Consequently, interference between the delayed copies of the scatter-broadened electric field manifests as scintillation. We measure a scintillation bandwidth \nud=$149\pm3$~Hz as well as a scatter-broadening timescale \taud=$1.22\pm0.09$~ms at 650~MHz. These two quantities are related through the uncertainty relation $C_1 = 2\pi$\nud\taud, using which we directly measure $C_1=1.2\pm0.1$. We describe the methods employed to obtain these results and discuss their implications in general, as well as for the magnetar XTE~J1810\textminus197, towards which the measurements have been made. We also discuss how such, effectively in-situ, measurements of $C_1$ can aid in inferring the wavenumber spectrum power law index and hence quantitatively discriminate between the various possible scattering scenarios in the ionized medium. Finally, using the fact $C_1 \sim 1$, we nominally constrain the emission size to less than a few 1000~km for a screen very close to the magnetar, and to within the magnetosphere for all screen distances.

\end{abstract}

\keywords{Radio transient sources (2008) --- Interstellar medium (847) --- Interstellar scintillation (855)}



\section{Introduction} \label{sec:intro}

Pulsars are observed to scintillate at radio frequencies. Scintillation arises from the interference of distorted wavefronts arriving at the observer as a result of multipath propagation through the ionized interstellar medium (IISM). Scintillation is typically observed as bright and faint patterns in the dynamic spectrum. 
Many nearby pulsars show striking scintillation patterns. For example, the nearby pulsar B1133+16 is now known to have at least six scattering screens \citep{McKee2022} that scintillate independently, while \cite{Ocker2024} find as many as nine screens to B1929+10 and four to B0355+54. J0437\textminus4715 is a nearby millisecond pulsar with 25 observed scattering screens \citep{Reardon2024} that are seen to scintillate individually, translating to an observed screen density 7$\times$-10$\times$ larger than previously known. The large number of scattering screens detected in nearby pulsars possibly suggests an abundance of such screens to distant pulsars with moderate and high DM.
In contrast, pulsars with higher DM, and hence at larger distances, appear to lack prominent scintles in their dynamic spectra. This does not imply that they do not scintillate, rather it is understood to be a limitation primarily arising from signal-to-noise, their small scintillation bandwidths (\nud) and limited spectral resolution available to resolve the scintles. However, given a sufficiently capable observing setup, it is possible to resolve scintillation, although quite possibly with diminished signal-to-noise. As long as the scintillation is intrinsically resolved by the instrument, it can be detected e.g. by performing an autocorrelation of the dynamic spectrum across the band. When the pulses are fully modulated, meaning that the fluctuations across the band are as strong as the signal itself, the scintillation can be measured if the pulse can be measured across the band.


The small \nud is understood to be a direct consequence of large scatter broadening time (\taud), due to the uncertainty relation  \citep{Cordes1998} (hereafter CR98),
\begin{equation}
2 \pi \Delta\nu_d \tau_d = C_1    
\label{eqn:uncertainty}
\end{equation}
 where $C_1$ is a constant determined by the waveneumber spectrum of the scattering material. This relation holds good for point sources, while the optics for extended sources, both intrinsic and scattered, are discussed in \cite{Gwinn1998}.
 
 Pulsars, owing to their steep spectra, are intrinsically brighter at frequencies where it is easier to measure \taud, while it is only a S/N limitation that often precludes a measurement of \nud\ at higher frequencies. The correlation between \taud and DM is empirically well-established \citep[e.g.][]{Mitra2001, Bhat2004}. This means that, often for moderate to high-DM pulsars with appreciable \taud, \nud\ is almost never measured directly; it is rather inferred with the aid of the uncertainty relation assuming a specific turbulence power spectrum (and hence $C_1$), often letting $C_1$=1. Even for low-DM pulsars with negligible scatter broadening and striking scintillation patterns which allows \nud to be robustly measured, observing with high spectral resolution can lead to interesting discoveries, such as with PSR B0834+06. The high spectral resolution VLBI observations and measurements reported by \cite{Brisken2010} led to the detection of a hitherto unknown second screen and associated exotic optical phenomena such as double lensing \citep{Liu2016, Zhu2023}. The 1-ms island, arising from refraction off the second screen, would not have been detected if not for the high delays probed in the secondary spectrum, aided of course by the baseband data acquired for VLBI correlations. 

For those pulsars that show appreciable \taud but measuring \nud is challenging, a direct measurement of both the quantities together could be attempted with Nyquist-resolution baseband data. Of course, the ability to measure both the quantities robustly depends on factors such as, e.g., an intelligent choice of the observing frequency band (which determines \nud\!\!), the S/N of the observed pulsed signal and the DM (which directly affects \taud\!\!). However, simultaneous measurement of both \taud and \nud can result in a direct, in-situ estimation of the parameter $C_1$. A direct estimate of $C_1$ can reveal the nature of the scattering material, as $C_1$ encodes the index $\beta$ of the wavenumber spectrum through the structure function index $\gamma$. Such a measurement of both \nud\ and \taud\ has been attempted earlier: e.g. by \cite{Walker2008} where they measure the scatter broadening time for B0834+06 from the sacttered wavefield obtained by deconvolving the secondary spectrum; by \cite{Main2017} from descattering B1957+20 giant pulses, for which they find \taud=12.2$\mu$s is consistent with the measured \nud; as well by \cite{Mahajan2023} similarly for B1937+21; by \cite{Walker2013} employing cyclic spectrscopy on B1937+21. In principle, $C_1$ can be estimated from independent measurements of \nud\ and \taud, which the above already obtain through different means.

In this work, we present our attempt to make an in-situ measurement of $C_1$ towards a particular source --- magnetar XTE~J1810\textminus197, which is currently active at radio frequencies. Of course, the magnetar nature of the source itself has no bearing upon the measurements, other than translating into implications for emission and scattering disc size. In addition to measuring $C_1$, this work is also motivated by the lack of any direct measurement of scintillation bandwidth towards this magnetar so far. Furthermore, intriguing spectral structures in the single pulses from this magnetar have been observed earlier \citep{Maan2019}. The origin of these spectral structures seemed likely to be intrinsic rather than scintillation, and a direct measurement of the scintillation bandwidth could potentially strengthen this deduction.

In the following sections, first we provide the relevant background information and further motivation for this work to be conducted on the particular target (\ref{sec:magnetar}) followed by details of the observing setup and data analysis (Section~\ref{sec:analysis}). We outline our results in Section~\ref{sec:results}, discuss their implications in Section~\ref{sec:discussion}, and summarize the conclusions in Section~6. Throughout this paper, we follow the convention  \nud$\propto\nu^\alpha$ and \taud$\propto\nu^{-\alpha}$, where $\alpha$ is a positive real number.

\section{Magnetar XTE J1810\textminus197} \label{sec:magnetar}
The magnetar XTE J1810\textminus197 ($18^\mathrm{h}09^\mathrm{m}51^\mathrm{s}.09$, $-19^\circ43'51.93''$) was discovered in X-ray \citep{Ibrahim2004}, with the distinction that it is the first transient Anomalous X-ray Pulsar (AXP). Subsequently, it was detected in radio \citep{Camilo2006} and observed to have highly variable radio pulsations at the same period as in X-ray, i.e., 5.54~s, and a remarkably flat radio spectrum $S \propto \nu^{-0.5}$ \citep{Camilo2007c}. Although strong and variable pulsations were seen in early observations, it was observed that the average pulse flux density was decreasing with time \citep{Camilo2007b} and eventually the magnetar became radio quiescent abruptly \citep{Camilo2016}. However, it has recently undergone an outburst in radio with a considerable increase in its average flux density \citep{LLS2018,Maan2019} at the outburst, followed by a secular decrease for many weeks. Additionally, more recent studies report a steeper and variable spectrum 
(see \citealt{MSJ2022} for details) 
after the late-2018 outburst. The source continues to be active, and hence, it is still an opportune time to study this intriguing magnetar in detail. 

Magnetar XTE~J1810\textminus197 has a DM of 178~\ppcc. It has a measured VLBI parallax distance of $\sim$2.5~kpc and a proper motion velocity of $\sim$200~km\ \!s$^{-1}$\citep{Ding2020}. To the best of our knowledge, there are no direct measurements of the scintillation bandwidth and timescale for this source. \cite{Maan2019} report spectral characteristics that resemble interstellar scintillation, but find that the characteristic bandwidths (visually, $\sim 20-40$~MHz) of these spectral features are inconsistent with the expected scintillation bandwidth derived from the NE2001 model, and hence conclude the features are likely intrinsic. \cite{Maan2019} also report a \taud $\sim 1.05-1.3$ ms ($\tau_\mathrm{sc}$ in their notation) at 650\,MHz, which translates into a small \nud$\sim 140-175$~Hz --- the only, indirect, measurement of \nud for this source. More recently, we have also measured a scattering timescale (unpublished) \taud $\sim 6$~ms at 400~MHz. Using a scaling index of $\alpha \sim 4$ for the frequency dependence, the expected scatter broadening at 650~MHz would then be \taud $\sim$~0.86~ms. This translates roughly into a \nud $\sim200$~Hz \citep[using $2\pi\Delta\nu_d\tau_\mathrm{d} = C_1$, and $C_1=1.16$;][]{NE2001, cordes+04} at 650~MHz, i.e., consistent with the estimates from \citet{Maan2019}.

Following the formulation in \citet{Pulsarhandbook2005}, \citet{Lazaridis2008} make an estimate of \nud $\sim$ 320~MHz and a scintillation timescale of 10~min at 15~GHz. Extrapolated with $\alpha\sim4$ to 650~MHz, the expected \nud $\sim 1.2$~kHz. This is $\sim 6\!-10\times$ larger compared to the expectation from the \taud measurement. However, this estimate could be off by a large amount and, additionally, a small deviation in $\alpha$ could easily explain the discrepancy between the two \nud values. Nevertheless, it remains to be seen if there is a second screen which, e.g., could have caused the spectral structures with typical bandwidths of 20-30\,MHz reported by \citet{Maan2019}. Moreover, measuring \nud at 650\,MHz also holds the potential for a direct, in-situ measurement of $C_1$ aided by the existing or simultaneous measurements of the scatter broadening time \taud.



\section{Observations and data analysis} \label{sec:analysis}
\subsection{Baseband voltage beamformer} \label{subsec:PASV}
XTE~J1810\textminus197 is being regularly monitored through an independent observing program. However, its scintillation bandwidth and timescale have never been directly measured. The scatter broadening timescale \taud reported by \cite{Maan2019} as well as unreported measurements at 400~MHz hint at the possibility of extremely small scintillation bandwidths $\lesssim 1$ kHz (at 1~GHz), requiring high-resolution spectroscopy. The phased array beamformer of the GMRT Wideband Backend (GWB) is limited by the GPU FLOPS and other bandwidth constraints to a maximum of 16384 channels. In principle, high spectral resolution can therefore be achieved by trading off observing bandwidth for band-integrated sensitivity. 

The ideal observing strategy would be to obtain baseband voltage data, so that near-arbitrary temporal and spectral resolution can be achieved. A newly-unlocked baseband beamformer mode in the GWB, primarily intended for VLBI observations, has therefore been used for the observations reported here. In the Total Intensity mode of the GWB, the beamformed complex voltage spectra are tapped from the output of the GWB F-engine and sent downstream for detection and integration after channelization using the fast fourier transform (FFT). In the newly emabled Phased Array Spectral Voltage (PASV) mode, the Nyquist-sampled data from the individual polarizations are recorded as complex spectra, i.e. the phased-array FFT output of the beamformer is collected from the ring-buffer shared memory and written to non-volatile memory (NVMe) disks, ensuring that the full 200/400 MHz of the baseband is captured without buffer loss. 
\subsection{Observing setup}
XTE J1810\textminus197 was observed with the upgraded GMRT (uGMRT; \citealt{uGMRT}) using the time awarded through a Director's Discretionary Time (DDT) proposal. Observations were conducted in three sessions over a total time of 4 hours. There were two single-band, one-hour observing sessions, one each at Band-4 (550-750 MHz) and Band-5 (1060-1260 MHz). In addition there was a two-hour long, Band-4/5 simultaneous dual-band observation with 100\,MHz bandwidth in each band. We recorded a single 200\,MHz PASV beam for the single band observations and 2 $\times$ 100\,MHz PASV beams for the dual-band observations with two disjoint sub-arrays comprising of different GMRT dishes. The spectrometer was configured with the default settings, giving 2048 channels across 100/200 MHz. The PASV  is a complex filterbank format, but it comes with associated header and timestamp files.



\subsection{Pre-processing}
The PASV filterbank data were first translated to DADA format and then converted to standard coherently dedispersed {\tt sigproc} filterbank using {\tt digifil}. For obtaining the dedispersed time sequence, the filterbank files were produced with 2048 channels with the best fit DM of 178.85~pc~cm$^{-3}$, returned by the {\tt pdmp} task of {\tt PSRCHIVE}. The time sequence was searched for bright single pulses using PRESTO \citep{PRESTO}. The detected pulses were then sorted in descending peak flux order and the brightest 20 pulses were considered for further analysis. Their peak-referred time of arrival (ToA) was used to identify the appropriate time segment in the PASV data. Short time segments, typically $\pm$2.5~s around the peak of the pulse, were then extracted from the PASV data, translated to DADA and {\tt sigproc} filterbank with extremely high spectral resolution.

The Band-4, 200\,MHz data were channelized into 2097152 (2M) channels, giving a spectral resolution of $\sim$ 95~Hz per channel. This is required for resolving the putative scintillation if it is expected to be smaller than 1~kHz. For the Band-5 100\,MHz data, the number of channels was set to 524288 (0.5M) resulting in spectral resolution half of that in Band-4. However, we use only the Band-4 data for our sensitive measurements as the Band-5 data turned out to be heavily contaminated by radio frequency interference (RFI) in the middle of the observing band. We therefore discuss the Band-5 data no further.

\begin{figure*}[t!]
    \centering
    \includegraphics[scale=0.6,trim=0.2cm 0cm 0cm 0.18cm, clip]{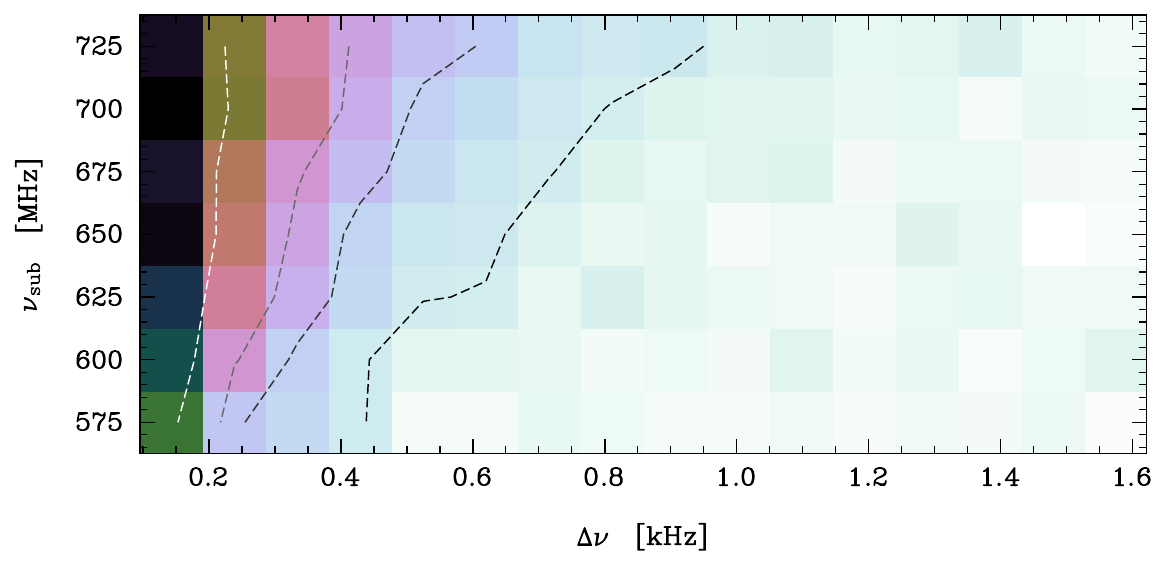}
    \includegraphics[scale=0.382,trim=0.0cm 0cm 0cm 0.22cm, clip]{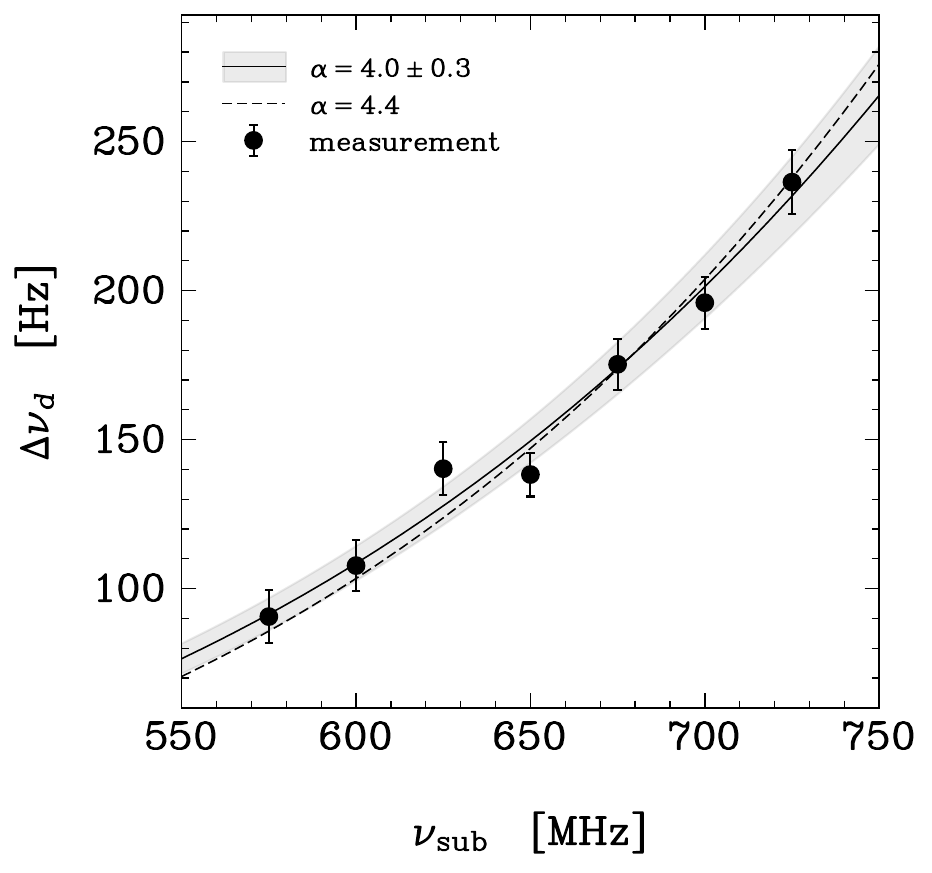}
    \caption{\emph{Left:} The subband-binned autocorrelation function stacked for 20 pulses vs frequency lag. \emph{Right:} The measured \nud values at each subband, along with the error bars and the fits.}
    \label{fig:scbw-measurement}
\end{figure*}

\subsection{Autocorrelation of single-pulse spectra}
The spectrum of each of the the identified pulses was obtained by integrating under the ON-gate\footnote{The ON-gate for each pulse was visually identified due to the complex morphology, but could be  defined as the contiguous signal between the extremal points of  the pulse roughly above $3.5\sigma$ of the root mean square off-pulse noise.} after weighting by the frequency-averaged profile. The spectrum thus obtained is subdivided into $M$ fine subbands, each consisting of $N$ channels. The autocorrelation function (ACF) is now calculated for each subband for $N-1$ frequency lags. The ACFs of the 20 identified pulses were stacked and averaged. For Band-4, $M=1024$ and $N=2048$. Finally, the 1024 subbands in Band-4 are grouped into 7 coarse subbands of equal frequency widths, after omitting the first and the last 64 fine subbands. Each coarse subband is thus 25-MHz wide. The final, frequency-resolved, ACF obtained is shown in Figure~\ref{fig:scbw-measurement}.

\subsection{Single pulse analysis for scatter broadening}

Every pulse emitted by the source undergoes a series of convolutions with multiple different response functions due to the propagation effects of the interstellar medium as well as the electrical response of the instrument. Broadly,
\begin{equation}
    S(t,\nu) = P(t,\nu) * G(t,\nu) * D(t,\nu) * I(t,\nu)
    \label{eqn:S_obs}
\end{equation}
where $S$ is the observed pulse signal, $P$ is the intrinsic pulse function, $G$ is the interstellar Green's function causing the scatter broadening, $D$ is the dispersion smear function and $I$ is the instrument response function \citep{Ramachandran1997}. Since the data were coherently dedispersed, the dispersion smear function $D$ is effectively a Dirac delta function; $I$ is  a $sinc$ function whose characteristic width is an inverse function of the frequency resolution set at the time of the observation.

\label{ssec:scattering-measurement}We assume a gaussian  for the intrinsic pulse shape and consider a one-sided decaying exponential for the pulse broadening function (PBF). In order to estimate the scatter broadening time, it is essential to identify bright, sharply peaked pulses with a simple morphology or isolated pulse components that can be fit with a single gaussian convolved with a decaying exponential. Pulses with complex morphologies require more gaussian components, resulting in typically poor fits with large off-diagonal covariance values. In the case of XTE J1810\textminus197, such as is the case for magnetars in general, a complex and unstable pulse profile evolution precludes easy identification of such sharply peaked single-component pulses. A thorough visual search resulted in one such reasonably bright pulse which, however, is not part of the main component. The baseband data segment for this pulse component was extracted from the larger file (as described earlier) and rechannelized to 256 channels to obtain a finer time resolution ($163.84~\mu$s) than that obtained for the scintillation analysis. 

\section{Results} \label{sec:results}
\subsection{Scintillation bandwidth} \label{ssec:scintillation-bandwidth}
We fit the ACF of each coarse subband in Figure~\ref{fig:scbw-measurement} with a decaying exponential $A e^{-\Delta\nu/\Delta\nu_d}$, excluding the zero-lag point, to measure the scintillation bandwidth \nud. The intervals between the isoclines are seen to increase with the subband frequency, as expected for scintillation. The fit parameters are listed in Table~\ref{table:scbw-fit}. 
The values returned by the fit at the coarse subbands show the expected power-law scaling behaviour with frequency. The measured values for \nud are then fit with a power law $\nu_d(\nu) = Q{(\nu/\nu_0)}^\alpha$, where $\nu_0$=650~MHz. The fit returned $Q = 149.5\pm3.3$~Hz and an index of $\alpha = 4.0 \pm 0.3$, which is also broadly consistent with $\alpha = 4.4$ expected for Kolomogorov turbulence. Such fine measurements are often challenging to realise, even more at lower frequencies, such as those by e.g. \cite{Wu2022}. 
The measured \nud\ allows us to independently confirm if the scatter broadening and scintillation arise from the same locales in the interstellar medium. In other words, independent measurements like these are necessary to investigate if the bulk of the scatter broadening arises from the local environs of the pulsar (magnetar in this case) while scintillation could be attributed to one or more intervening screens. Often, for lack of an independent measurement, the inferred scintillation bandwidth (through the uncertainty relation) is attributed to the scatter broadening.

\begin{table}[]
    \centering
    \begin{tabular}{c|c|c}
\hline
\hline
 Frequency  & A & \nud \\
  MHz  &                &    Hz\\
\hline
575 & 1.45 $\pm$ 0.18 & \ 90.6 $\pm$ \ \!8.8 \\
600 & 1.47 $\pm$ 0.13 & 107.7 $\pm$ \ \!8.6 \\
625 & 1.23 $\pm$ 0.08 & 140.2 $\pm$ \ \!8.8 \\
650 & 1.54 $\pm$ 0.08 & 138.3 $\pm$ \ \!7.4 \\
675 & 1.24 $\pm$ 0.06 & 175.3 $\pm$ \ \!8.6 \\
700 & 1.30 $\pm$ 0.05 & 195.9 $\pm$ \ \!8.8 \\
725 & 1.07 $\pm$ 0.04 & 236.4 $\pm$ \!11.7 \\
\hline
    \end{tabular}
    \caption{The measured parameters of the one-sided exponential that fit the autocorrelation function giving the \nud values for each subband and the respective error bars.}
    \label{table:scbw-fit}
\end{table}

\subsection{Scatter broadening time}
\begin{figure*}[t!]
     %
    \includegraphics[trim=0.0cm -0.90cm 0cm 0.0cm, clip, width=0.85\linewidth]{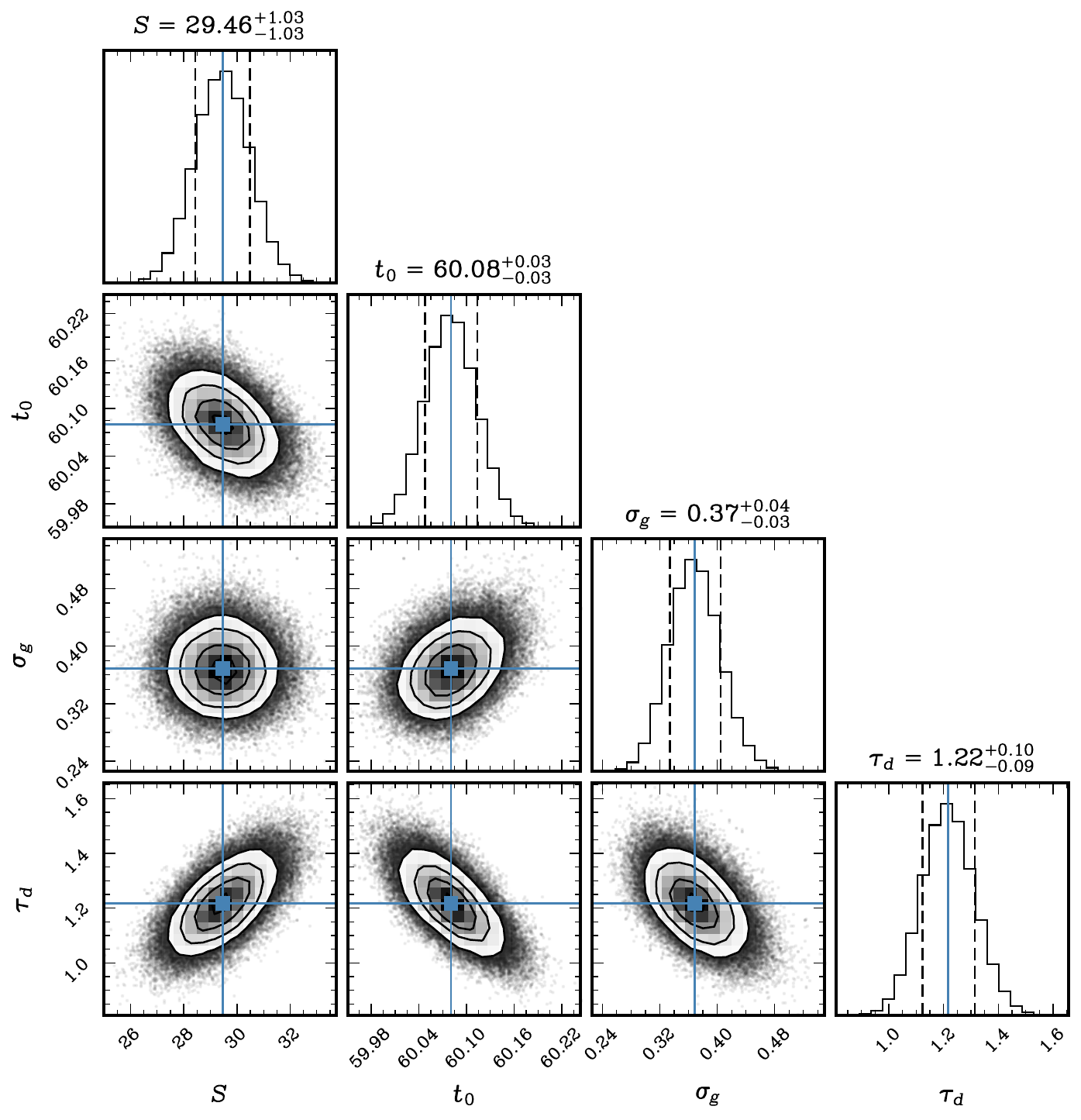} 
    \hspace{2.3cm}
    \llap{\raisebox{10.0cm}{
    \includegraphics[height=6cm]{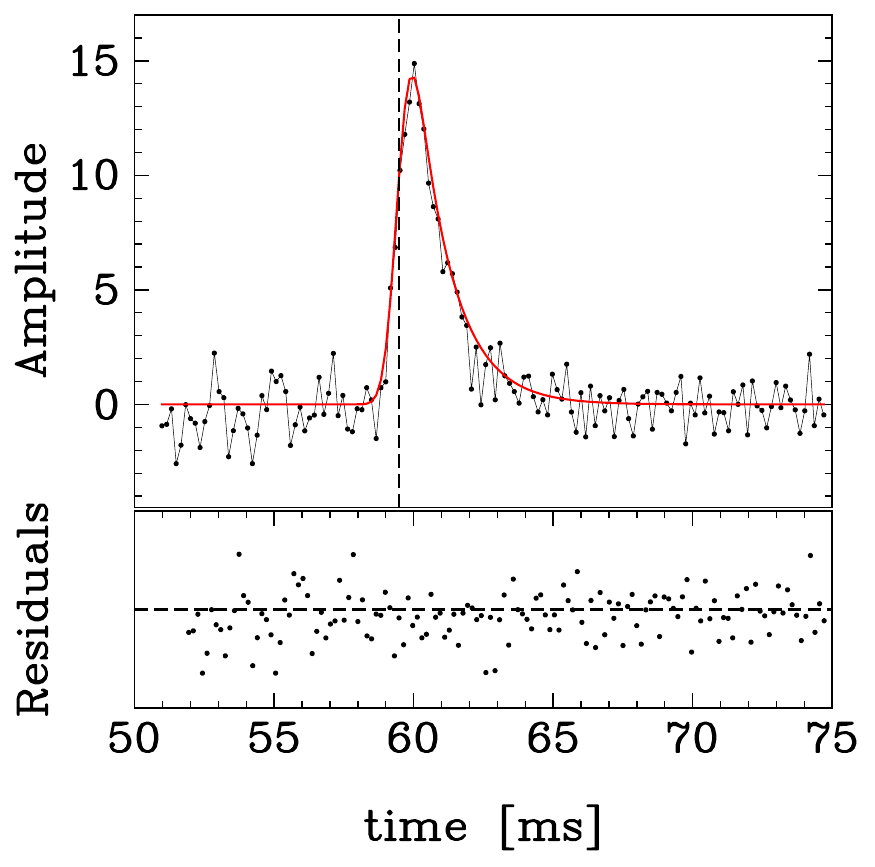}
    }}
    \caption{The scatter-broadened pulse component, the fit and the residuals, along with the posterior distribution of the fit parameters.}
    \label{fig:tau}
\end{figure*}

Figure~\ref{fig:tau} shows the isolated pulse component used for fitting the scatter broadening time. We fit a gaussian pulse $p(t)$ convolved with a decaying exponential $g(t)$ (which is the PBF) to the data, 
\begin{equation}
    s(t) = p(t) \ast g(t) \\
     = \frac{S}{\sigma_g \sqrt{2\pi}}\ e^{- \frac{1}{2}\left( \frac{t-t_0}{\sigma_g} \right) ^2} \ast\  \frac{1}{\tau_d}\ e^{- \frac{t}{\tau_d}}
    \label{eqn:scat-model}
\end{equation}
where $t_0$ is the instant at which the gaussian peaks, $\sigma_g$ is the standard deviation and $S$ is the fluence of the scatter-broadened pulse.
While fitting a single broadened pulse profile over a wide band, care has to be exercised as the \taud\ scales according to a power law across frequency. We apply a correction to the monochromatic frequency $f_m$, considering the central frequency $f_c$ of the observing band with a bandwidth $\delta f$, following equation~6 of \cite{Geyer2016}. We obtain $f_m = 657.65$~MHz for $f_c=650$~MHz over 200 MHz: the measured scatter broadening time at 657.65~MHz is \taud\!$=1.22^{+0.10}_{-0.09}$~ms which scales to $228^{+18}_{-16}$~$\mu$s at 1~GHz for $\alpha=4.0$. The inverse of the spectral channel width makes an additional contribution to the scatter broadening time through a \emph{sinc} function, which we ignore as it is $\sim 1.3~\mu$s for 256 channels across 200~MHz. Effectively, we approximate the \emph{sinc} function as a $\delta$-function for $I(t,\nu)$ in equation~\ref{eqn:S_obs} and it is small enough to be subsumed under the estimated $\pm8\%$ error. 

\section{Discussion} \label{sec:discussion}
\subsection{Scintillation}

Detecting and measuring scintillation is essentially limited by spectral resolution and S/N, assuming that the locales where scintillation arise are the same as those responsible for the scatter-broadening, i.e., they arise from the same ``screen''. This is largely true, except in those cases where the scatter-broadening may result from a dynamically active region closer to the source, whereas the scintillation may arise in a second screen: if the screen closer to the observer still ``sees'' an unresolved source despite scatter-broadening, it would cause scintillation independent of the scattering screen. This seems to be somewhat more common in fast radio bursts, although not surprisingly, where the scatter broadening typically arises in the host galaxy whereas the Galactic ISM ``sees'' an essentially unresolved source \citep[see][for examples]{Masui2015, Main2022, Marthi2022, Sammons2023}. In contrast, e.g. in the Crab pulsar \citep{Main2021}, deviation from the uncertainty relation results from a nebular screen resolving the scattering closer to the pulsar and not the result of a two-screen effect. 

\cite{Lazaridis2008} note that at the frequency of their observations, i.e., at 14~GHz, the flux density variations are expected to be large, attributable to scintillation. They argue that the large observing bandwidths and long integration times would average many scintles, thereby yielding reliable flux density estimates. They consider a critical frequency $\nu_c$=14.6~GHz, above which scintillation is thought to transition from strong to weak. This $\nu_c$ is 
an estimate based on earlier measurements \citep{Malofeev1996} and the NE2001 electron density model \citep{NE2001}. They therefore estimate, using expressions given in \citet{Pulsarhandbook2005}, a diffractive scintillation bandwidth \nud$=320$~MHz and a diffractive scintillation timescale $t_d=600$~s. 


The fit to the measurements (see Figure~\ref{fig:scbw-measurement}) returns $\Delta\nu_d(650~\mathrm{MHz}) = 149\pm3~\mathrm{Hz}$ and $\alpha=4.0\pm0.3$, which is also consistent to within 2.5\% with a forced Kolmogorov $\alpha=4.4$ that has been locally fit through the measurements. The shaded  region is the $\pm 1\sigma$ error bar for the fit obtained by formally propagating the errors on the $\alpha$ and the \nud measurements. Our \nud measurement at 650~MHz, when extrapolated to 14~GHz is $\sim 32$~MHz, an order of magnitude smaller.

\cite{Maan2019} detected spectral features which, being inconsistent with the scintillation bandwidth estimated from \cite{NE2001} as well as that deduced from their own scatter-broadening measurements, they concluded as not arising due to scintillation. Our direct measurements of the scintillation bandwidth also support the above conclusion. Furthermore, \cite{Gothelf2004} present X-ray observations that indicate an absence of any plerionic material around XTE J1810\textminus197. Nonetheless, whether these features are intrinsic and linked to the outburst itself or whether linked to an influence of the outburst on the propagation in the vicinity of the magnetar needs further investigation. If the features are phenomenologically linked to the outburst, we would expect them to also fade with time. High cadence monitoring observations after an outburst could help shed some light on the nature and possible origin of the spectral features.

No \HII\ emission is detected along the LoS to this magnetar. Consequently, there is no observational signature suggesting the presence of an additional screen which could have caused the observed spectral features. The NE2001 model underestimates the DM ($\sim$115\ppcc) to the parallax-measured distance of $2.5^{+0.4}_{-0.3}$~kpc \citep{Ding2020} while, of course, it overestimates the distance ($\sim$3.35~kpc) for DM=178 \ppcc. Small scale LoS variability of the DM may not be adequately modeled by NE2001, and so such discrepancies are not very surprising. In this context, \cite{Ocker2024} note that for pulsars with significantly higher DMs, especially a large fraction of those with extremely high \taud\ ($>$ 10~ms at 1~GHz), intersect \HII\ regions. Since such pulsars are often those with DM $>$ 600 \ppcc\ and significantly underestimated by NE2001, our non-detection of LoS \HII\ is consistent with their findings.

In addition to the scintillation bandwidth \nud, we attempted also to measure the scintillation timescale $t_\mathrm{d}$. This was done by picking adjacent pairs of pulses, with one of each pair coming from the selection of the 20 bright pulses, and computing the cross-correlation function in the same way as the auto-correlation function, but no cross-correlation was detected. \cite{Lazaridis2008} estimate a scintillation time $t_\mathrm{d}\sim 600$~s at 15~GHz, which scales to 26~s at 650~MHz. The Galactic Electron Density model NE2001 \citep{NE2001} predicts $t_\mathrm{d}\sim12.5$~s for this line of sight for DM=178.85~pc~cm$^{-3}$ at 1~GHz, which translates to $6.9$~s to $9.4$~s in the GMRT Band-4. The absence of cross-correlation is puzzling, and any definitive claim on $t_\mathrm{d}$ is deferred until further analysis, possibly at a higher frequency.

\subsection{Scatter broadening} \label{ssec:scatter-broadening}
Figure~\ref{fig:tau} (corner plot) shows the posterior distribution of the maximum-likelihood estimates of the parameters $S$, $t_0$, $\sigma_g$ and \taud, obtained through a Markov Chain Monte Carlo (MCMC) simulation. The best fit model and the residuals are shown in the top right panel. We note that, and not surprisingly, that $\sigma_g$ and \taud are a mildly degenerate pair. $S$ is uncorrelated with $\sigma_g$ since it merely scales the normalized gaussian. Ideally, one would expect $S$ to be uncorrelated with \taud, but note that there is a $S$/\taud in equation~\ref{eqn:scat-model} that could explain the partial correlation.

The \taud\ measured is the characteristic time of a decaying exponential, with which a gaussian is convolved: while we have used \taud\ in the uncertainty relation given by equation~\ref{eqn:uncertainty}, the more appropriate quantity to use is the mean delay time $\tau_\mathrm{delay}$ (see eqn. 8 of CR98). For a thin screen ($\Delta s/s \ll 0.01$), $\tau_\mathrm{delay}$=\taud, since the scattered pulse can be considered the result of a convolution of the intrinsic pulse with a single decaying exponential. For a thick screen, the mean delay time $\tau_\mathrm{delay}$ is not the same as the scatter broadening time \taud, and cannot be measured unambiguously. CR98 define \taud\ as an infinite sum of decay times $\tau_{d_n}$, where each contribution comes from an infinitesimally thin screen. Since recursive convolution of decaying exponentials eventually produces a gaussian, it is possible to arrange for a large number of screens such that the dominant screen makes the largest contribution to scatter broadening, while the gaussian ``intrinsic pulse''  is degenerate with multiple sub-dominant instances of scatter broadening. The measured \taud is hence the most dominant contribution to the effect.

We consider only the simplest PBF, which is the decaying exponential. We have a single scatter-broadened pulse component in our data to deduce the decay time. In principle, we could fit other PBFs, such as those listed in \cite{Bhat2004}, \cite{Lambert1999} and \cite{NE2001}. 
In passing, we note that giant pulses, which are on average much narrower than the mean pulse profile, allow us to measure \taud\ much more precisely, where possible: e.g. the black widow pulsar B1957+20 \citep{main+18a} and the millisecond pulsar B1937+21 \citep{Kinkhabwala2000}.
An unstable and very wide (compared to the scatter broadening) pulse profile precludes such analysis robustly for the magnetar J1810\textminus197.

We also measure a modulation index $\rho\sim0.71$ for all the pulses considered in this work, and find that it is invariant across the band. This is important since a varying modulation across the band would suggest that the source appears extended to the scattering screen, which can happen when the scattering screen is close enough to the emission region. While it would be instructive to measure the variation of $\rho$ across the scattering tail, the extremely fine \nud\ does not afford us the luxury to sample the scattered pulse at sufficient time resolution. A scatter-broadened profile with discrete features poorly fit by an exponential would suggest a turbulence power spectrum ill-parametrized by a power law. The post-fit residuals of our measurements in Figure~\ref{fig:tau} resulting from a reasonable fit of the model to the data, suggest otherwise.

\subsection{$C_1$}
Given that the effective frequency at which \taud\ is measured is corrected upwards, we obtain \nud=156~Hz at $f_m=657.65$~MHz from the powerlaw fit obtained in Section~\ref{ssec:scintillation-bandwidth}.
The uncertainty relation now gives $C_1 = 1.2\pm0.1$, where the 1-$\sigma$ error bar is computed from the respective errors in fits to the measurements of \nud and \taud. We denote the structure function index by $\gamma$ to avoid confusion with the power law index for frequency scaling $\alpha$ (which is the structure function index in the notation of CR98). Our estimate of $C_1$ is consistent with a uniform medium with Kolmogorov wavenumber spectrum ($\gamma=5/3,\  C_1 = 1.16$). However, it is also in agreement with with a thin screen with structure function index $\gamma=2$ ($C_1 = 1$) at $2\sigma$, as well as a thin screen with a structure function index $\gamma = 5/3$ (i.e. Kolmogorov; $C_1 = 0.957$), at 2.4$\sigma$. Our measurement rules out a Kolmogorov screen with a square law structure function ($\gamma=2,\ C_1=1.53$) at $>3\sigma$.


\begin{figure*}[t!]
    \centering
    \includegraphics[scale=0.75, trim=0.0cm 0cm 0cm 0.2cm, clip]{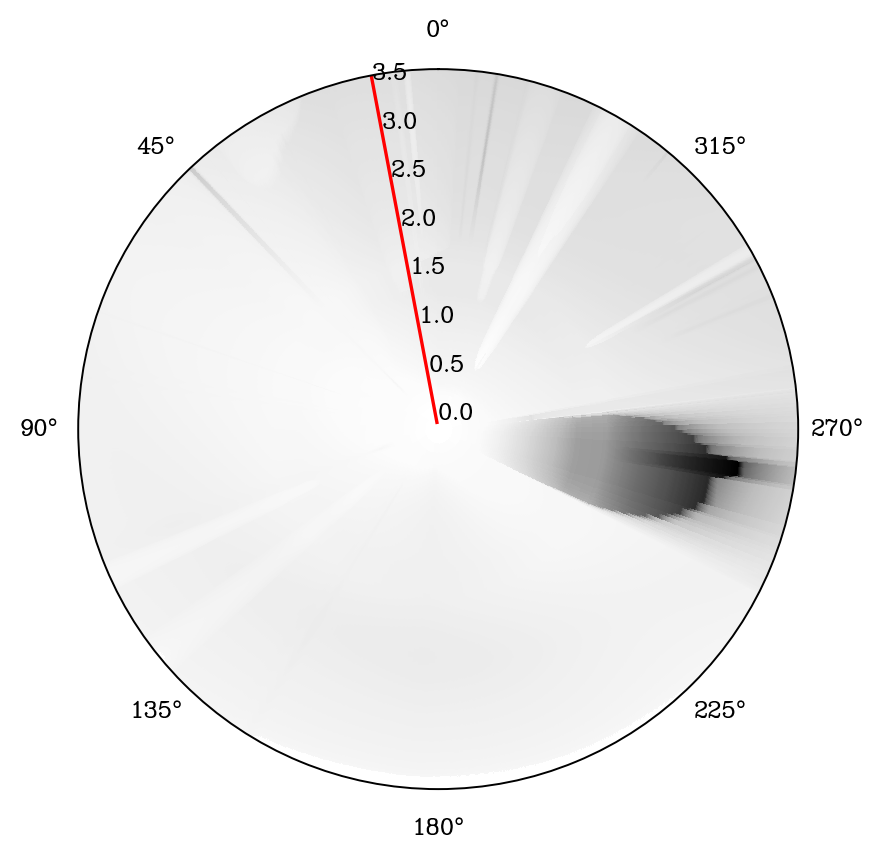}
    \includegraphics[scale=0.69, trim=0.0cm 0cm 0cm -0.52cm, clip]{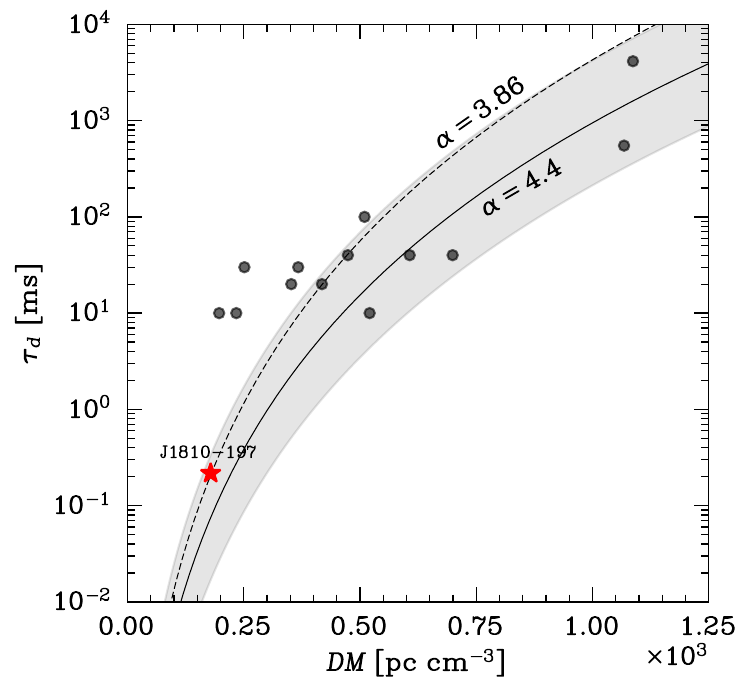}
    \caption{\emph{Left:} The electron density model for $gb=-0.158^{\circ}$, the Galactic latitude of J1810\textminus197, out to a distance of 3.6~kpc from Earth. The red marker line is in the direction of the magnetar, with the distances marked in kpc. \emph{Right:} The DM-\taud of \cite{Bhat2004} is represented by the dashed line; the solid line and the shaded region represent the empirical fit and the error bar $\sigma_{\tau_d}$ with $\alpha=4.4$ obtained from \cite{NE2003}, all at 1~GHz. The points represent pulsars within $10^\circ$ of XTE J1810\textminus197 beyond 2.5 kpc; the Galactic center lies just outside the $10^\circ$ circle.}
    \label{fig:NE2001}
\end{figure*}

\subsection{Implications for the scattering medium}
The index $\alpha$ is determined by the wavenumber spectrum in the medium through a power law (with index $\beta$; e.g. \citealt{Rickett1977}, \citealt{Krishnakumar2015}) of the power spectrum of the electron density fluctuations. Being a LoS-integrated measure, local conditions along the LoS cannot be inferred. \cite{Bhat2004} note that departure from the frequency scaling index $\alpha=4.4$ is possible when there is an inner-scale cut-off in the wavenumber, or an intrinsically non-Kolmogorov wavenumber spectrum. The third effect they consider is bulk refraction that could bias the scatter broadening time upward. However, $C_1 \sim 1$ rules out the particular two-screen model where the scatter-broadened magnetar appears unresolved to a foreground screen.

\cite{Ding2020}, besides measuring a parallax distance, also measure the size of angular broadening of 0.7$\pm$0.4~mas at 5.7~GHz. Additionally, they note the presence of a jet-like feature of $\sim$10~mas, substantially resolved on the longest VLBA baselines. The angular broadening scales to 54~mas ($\theta \propto \lambda^2$) at 650~MHz. Note that the light cylinder radius is $2.65\times 10^5$~km for the magnetar, suggesting potentially large emission heights. Thus, the measured modulation index of $\rho=0.71$ assumes significance. VLBI observations of the scattering at lower frequencies (relative to that at which the parallax is measured, say 1.4~GHz) would be very interesting and useful to localize the scattering material along the line of sight.

Figure~\ref{fig:NE2001} shows the Galactic NE2001 model \citep{NE2001} radial electron density profile for the slice through $gb = -0.158^\circ$ for $0^\circ < gl \leq 360^\circ$, indicating no significant density enhancement along the direction of the magnetar out to a distance of 3.5~kpc. However, due to its negative latitude, the LoS to the magnetar from Earth (which is $\sim 30$~pc above the Galactic plane) encounters a significant amount of material in the plane. The empirical DM$-$\taud relation along with the $\pm 1\sigma$ error is shown in the right panel. The black points indicate pulsars $>$ 2.5~kpc around a $10^\circ$ radius of the magnetar for which \taud is measured. A few pulsars show significant departure above $1\sigma$, but this is not entirely unexpected --- these could be reconciled by invoking a shallow index for $\alpha$ for particular lines of sight, e.g., as a result of upscattering due to bulk refraction. Alternatively, the scattering of these pulsars is likely not described well by powerlaws, or described reasonably by broken powerlaws for finite screens.


The observations reported here cannot discriminate reliably between the two possiblities, namely between an inner-scale cutoff and non-Kolmogorov spectrum, for the line of sight to XTE J1810\textminus197.  The error bars on the measured \nud and \taud are limited by (a) the inability to utilise the full exposure on the magnetar to estimate \nud, as only the brightest 20 pulses were considered due to the unstable pulse morphology and (b) estimating \taud from a single pulse component due to the complex, multi-component profile. A similar analysis may be attempted for a sample of bright pulsars where it may be possible to measure both \nud from the full scintillation dynamic spectrum and \taud from the PBF much more reliably, gaining from the stable folded profile. In fact, millisecond pulsars with measurable scatter broadening time of \taud$\sim1$~ms can serve as excellent probes, for which one can also measure \nud as described above. Measured over multiple independent sightlines to different millisecond pulsars with a range of DMs, one might be able to get more accurate estimates of $C_1$ across a subset of the observable (DM, \taud\!\!) parameter space.
\subsection{Constraints on the size of emission}
Since scintillation arises from interference of coherent beams of radiation arriving through different paths, it has the intrinsic ability to resolve angular detail. The modulation index $m$ is a qualitative predictor for the source size (e.g. \citealt{Gwinn1998}, \citealt{Johnson2012}). For fully modulated scintillation $m\sim 1$, while $m \lesssim 0.2-0.3$ typically indicates weak scintillation. Intermediate indices $0.5<m<0.8$ qualitatively suggest a range of possibilities between well-resolved to barely resolved.

The measured \taud$\sim1$~ms translates into an angular broadening \citep[e.g.][]{Simard2019a} of 
\begin{equation}
    \theta_d = \sqrt{2c\tau_d\frac{d_p - d_s}{d_pd_s}}
\end{equation}
where $d_p$ and $d_s$ are the distances to the magnetar and the screen from the observer.
This scattering results in an intrinsic angular resolution at the observing wavelength $\lambda$, given by
\begin{equation}
    \theta_\mathrm{res} = \frac{\lambda}{D}
\end{equation}
where $D$ is the lateral extent of the scattering at the screen, given by $D = \theta_d d_s$. The scattering hence defines a resolution element at the magnetar, which is the lateral extent of the emission region over which the phase at the screen changes by $\sim 1$~rad, given by
\begin{equation}
    r_\mathrm{res} = \theta_\mathrm{res} (d_p - d_s)
\end{equation}
Substituting for $\theta_\mathrm{res}$ gives
\begin{equation}
    r_\mathrm{res} = \lambda \sqrt{\frac{1}{2c\tau_d}\frac{(d_p - d_s)d_p}{d_s} }
\end{equation}

Since the screen is not localized, we first consider the limiting case of the screen practically at the source in relation to its distance from the observer, e.g. $d_p - d_s=1$~pc. Note that this is of similar order in scale as the Crab nebula, whose size is $\sim 3.5$~pc at a distance of 2~kpc. The 1~pc screen places an upper limit on the emission region size of $r_\mathrm{res} \lesssim 100$~km. The light cylinder diameter is $2r_l\sim 5\times10^5$~km, while the radio emission from $\sim1000$~km above the polar cap region has a much smaller size. Several studies suggest that the emission diameter is $\sim 1000-2000$~km \citep{Gwinn2012, RLin2023}, even as there is no evidence for evolution of the emission height with period \citep[see e.g.][Figure~2]{Mitra2023}. The measured modulation index $m=0.71$ suggests a partially or barely resolved emission region, which conservatively constrains the emission size to a few resolution elements, i.e. well within the light cylinder. This is consistent with expectation, since the uncertainty relation (equation~\ref{eqn:uncertainty}) holds for a point-like source, and the measured $C_1 \sim 1$. 
\begin{figure}
    \centering
    \includegraphics[width=1.0\linewidth]{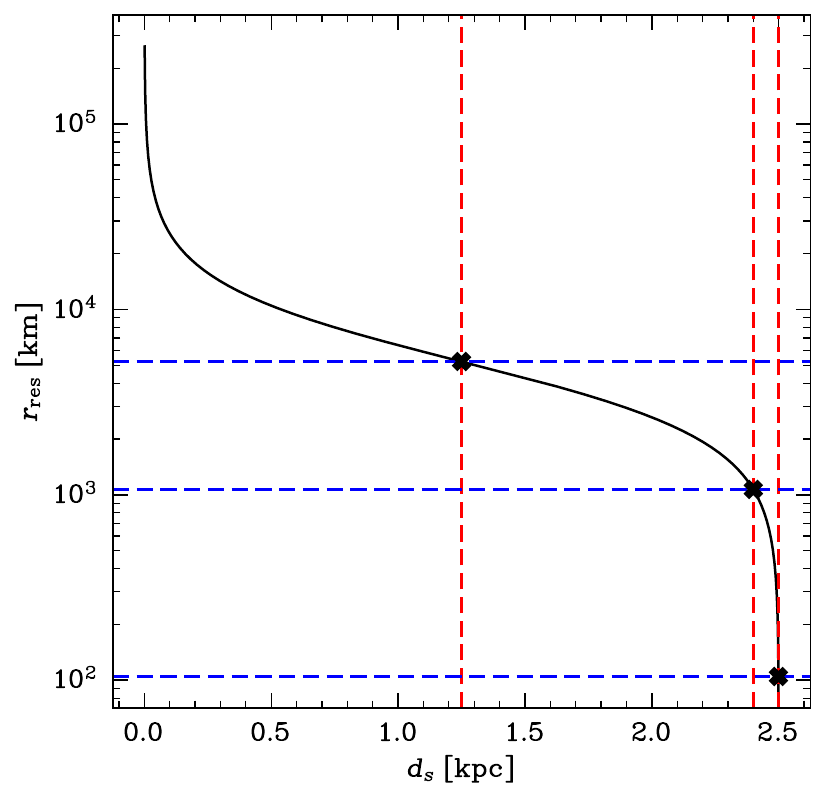}
    \caption{The emission region size at the magnetar constrained by the scattering on a screen at a distance $d_s$ from the observer. The crosses mark the coordinates $(d_s, r_\mathrm{res})$ for screens at 1~pc, 100~pc and half-way from the magnetar.}
    \label{fig:emission-size}
\end{figure}
Screens further from the magnetar impose less stringent requirements on the emission size: for instance, a screen 100~pc from the magnetar would diminish the resolving power by an order of magnitude to $r_\mathrm{res} \lesssim 1000$~km. Figure~\ref{fig:emission-size} shows the evolution of the resolution element as a function of screen distance for the masured \taud$\sim1$~ms. This is important in the context of fast radio bursts, for which magnetars are a leading candidate for the progenitor \citep{Bochenek2020}. Several FRBs have been shown to have an associated host-galaxy screen \citep{Masui2015, Main2022, Nimmo2025}. Because of the long period of magnetar J1810\textminus197, Figure~\ref{fig:emission-size} shows that the emission can be constrained to be magnetospheric (i.e. $r_\mathrm{res} < 2r_l$) for all screen distances. This is generally not true for pulsars, where only screens closer to the source are able to effectively constrain emission as magnetospheric, given their typically smaller rotational periods. Our result is also consistent with the recent direct constraint on FRB emission size to $\sim10^4$~km, suggesting a magnetospheric origin \citep{Nimmo2025}. In the instant case, the stringent constraint on the size of emission $r_\mathrm{res} < 1000$~km is aided by the large \taud, while additionally the relaxed magnetospheric upper limit of $r_\mathrm{res} < 2r_l$ for all screen distances comes from the magnetar's $P_0 \sim5.5$~s.

For an unresolved source, the resolution element places an upper limit on the size of the emission region. Observationally, this manifests as full modulation of the pulses, i.e. $m=1$. The reduced modulation index measured in our observations suggests that the emission region is partially resolved, which still constrains the size to well within the magnetosphere given its modest value $m \sim 0.71$. Note that since temporal resolution is compromised for the high spectral resolution required for measuring the scintillation in this work, the modulation index is averaged over the entire pulse. Progressive dilution of the modulation in the scattering tail, if observed, would place the tightest constraint on the emission size. The intrinsic limitation imposed by the fine scintillation rules out such an analysis with our observations.


In passing, we reiterate from \ref{ssec:scatter-broadening} that it is possible to arrange a succession of thin screens to explain the observed \taud, in which case the details of the screen configuration would determine the observed intensity modulations. Such an analysis might benefit from additional observational aids such as VLBI.


\section{Summary}
Magnetar XTE J1810\textminus197 is among the handful that continues to be active in radio frequencies, well past its last outburst in X-ray and hence being continually monitored for spectral and flux evolution \citep{Maan2019}. Although some spectral features resembling scintillation were observed, these have been concluded to have been intrinsic, as the measured characteristic spectral widths are not commensurate with the scatter broadening times. Our \nud measurement from this work further strengthens the above conclusion. Though, it is not yet clear if these features were related to the 2018 outburst. 

We have reported the first ever direct measurement of the scintillation bandwidth \nud\ for this magnetar at any frequency, simultaneously with \taud. We stack the autocorrelation function of the brightest pulses to estimate \nud$=149\pm3$~Hz at 650~MHz. Along with \taud$=1.22\pm0.09$~ms, these measurements enable us to directly infer $C_1=1.14\pm0.09$, which depends on the power-law index $\beta$ of the electron density wavenumber spectrum and hence the index $\alpha$ of the frequency dependence of the scatter broadening timescale \taud. While our measurement of $C_1$ rules out a two-screen model, it is unable to robustly discriminate between Kolmogorov ($\gamma = 5/3$), as well as between a uniform scattering medium and a thin screen. A uniform scattering medium with a square law phase structure function is also ruled out at 4$\sigma$. The extended IISM can be represented as a succession of randomly oriented, infinitesimally thin screens, such that the screen with the most dominant contribution to the scatter broadening leads to the measured \taud, while the multitude of subdominant screens convolve out an arbitrary intrinsic pulse function to a gaussian.

Following the successful demonstration of the feasibility of making direct measurements of $C_1$ in this work, we propose systematic, baseband observations of moderate- and high-DM  pulsars along different sightlines to build a sample of the LoS IISM out to different DMs and distances. Specifically, millisecond pulsars with extremely stable pulse profiles and those with giant pulse emission are excellent candidates for cyclic spectroscopy. Empirical estimation of  $C_1$ along multiple sightlines will aid in refining the electron density tomograph of the Milky Way.

Finally, we consider constraints on the size of the emission for nominal screen distances. A single scattering screen at any distance constrains the emission to within the magnetosphere since the resolution element of the screen is smaller than $2r_l$, which is consistent with the uncertainty relation holding for point-like sources. This constraint assumes importance within the context of a growing body of work that posits magnetars as a progenitor for fast radio bursts.

\section{Acknowledgements}
We thank the reviewer for the detailed comments on the manuscript, which have helped substantially improve the content and presentation. We thank the staff of GMRT who made these observations possible. GMRT is run by the National Centre for Radio Astrophysics of the Tata Institute of Fundamental Research. We gratefully acknowledge the Department of Atomic Energy, Government of India, for its assistance under project No. 12-R\&D-TFR-5.02-0700. YM acknowledges support from the Department of Science and Technology via the Science and Engineering Research Board Startup Research Grant (SRG/2023/002657). VM acknowledges the SciNet HPC Consortium. Some computations were performed on the Niagara supercomputer at the SciNet HPC Consortium. SciNet is funded by Innovation, Science and Economic Development Canada; the Digital Research Alliance of Canada; the Ontario Research Fund: Research Excellence; and the University of Toronto.


%

\facilities{Giant Metrewave Radio Telescope (GMRT)}


\software{\tt dspsr}\citep{dspsr}, {\tt RFIClean} \citep{MLV21}, {\tt NumPy} \citep{numpy}, {\tt astropy} \citep{astropy:2022}, {\tt matplotlib} \citep{matplotlib}, {\tt smplotlib} \citep{smplotlib}, {\tt statsmodels} \citep{statsmodels}, {\tt emcee} \citep{MCMC}, {\tt corner} \citep{corner}.






\newpage
\bibliography{XTE}{}
\bibliographystyle{aasjournal}



\end{document}